\documentclass{article}

\usepackage[T1]{fontenc}
\usepackage[utf8]{inputenc}
\usepackage{bm}
\usepackage{dsfont}
\usepackage{graphicx}

\usepackage[qm]{qcircuit}

\newtheorem{defn}{Definition}

\renewcommand{\vec}[1]{\bm{#1}}

\newcommand{\ket}[1]{\bm\vert\,{#1}\,\bm\rangle}

\newcommand{\set}[1]{\{#1\}}

\newcommand{\booldomain}[1][n]{{\set{0,1}}^{#1}}

\newcommand{\tensorpower}[2][n]{{#2}^{\otimes{#1}}}
\newcommand{\bigo}[1]{\mathcal{O}\bigl(#1\bigr)}

\newcommand{\vecx}{\vec{x}}
\newcommand{\vecp}{\vec{p}}
\newcommand{\Hilbert}{\mathcal{H}}

\usepackage{authblk}
\title{On the connection between Deutsch-Jozsa algorithm and bent functions}
\author[1]{ERALDO PEREIRA MARINHO}
\affil[1]{Univ Estadual Paulista -- UNESP\\
          Department of Statistics, Applied Mathematics and Computing\\
    	  Avenida 24 A, 1515, DEMAC\\
	  Rio Claro, Sao Paulo, ZIP 13506-900, Brazil\\
          email:pereira.marinho@unesp.br, ORCID: 0000-0002-7413-0161}

\begin{document}
\maketitle

\begin{abstract}
It is presently shown that the Deutsch-Jozsa algorithm is connected to the
concept of bent function. Particularly, it is noticeable that the quantum
circuit used to denote the well known quantum algorithm is by itself the quantum
computer which performs the Walsh transform of a Boolean function.
Consequently, the output from Deutsch-Jozsa algorithm, when the hidden function
is bent, corresponds
to a flat spectrum of quantum states.
\end{abstract}

\section{Introduction}
The concept of bent function was originally introduced by \cite{Rothaus:1976},
for cryptography purposes, and is intuitively the Boolean function that most
distances from linear, $f(\vecx)=\vec k\cdot\vecx$, and affine, $f(\vecx)=\vec
k\cdot\vecx\oplus 1$ cases. The dot-product notation corresponds to the
Boolean inner product, namely
\begin{equation}
 \vec{k}\cdot\vecx=
 \sum_{j=0}^{n-1}k_jx_j
 \mathtt{mod}2
 \equiv
 \bigoplus_{j=0}^{n-1}k_jx_j
\end{equation}

\par
What is interesting, and is shown in this article, is that the general
form of the Deutsch-Jozsa \cite{Deutsch-Jozsa:1992} algorithm coincides with
the Walsh transform. This in turn has constant absolute value in the case of
bent functions. Consequently, the famous quantum algorithm reproduces a flat
spectrum in its output if the hidden function is a bent function; which is not
the aim of the original application for the algorithm
\cite{Deutsch-Jozsa:1992}. Thus, such peculiarity of the Deutsch--Jozsa
algorithm provides a fast $\bigo1$ check if an unknown function is or not a
bent function.

\par
In Section 2, we will briefly summarize the Walsh transform on $n$-bit Boolean
functions. In Section 3, it will be formally shown that the definition of bent
functions reproduces a flat response spectrum of the Deutsch--Jozsa algorithm.
In Section 4, we will show some examples obtained from a simple classic
simulator of the famous quantum algorithm, for the bent and non-bent case of
Boolean functions over a 4-bit space. Some conclusions and perspectives will be
discussed in Section 5.

\section{The Deutsch--Jozsa algorithm and the Walsh transform of a Boolean
function}
\subsection{The Deutsch--Jozsa algorithm}
\par
To recall, if the test function $f:\booldomain\mapsto\booldomain[]$ is constant,
either $f=0$ or $f=1$, the Deutsch--Jozsa algorithm responds with the
monochromatic
state $\ket{\psi}=\ket{0^n}\equiv\tensorpower{\ket{0}}$.
On the other hand, if the output state is $\ket{\psi}\ne\ket{0^n}$, the
algorithm is
indicating that the function is balanced. However, such conclusion is a question
of
faith once it requires the belief that the function has been programmed to
behave
as a constant otherwise as a Boolean balanced function.

Particularly, \cite{Marinho:2018} pointed out that the algorithm
responds with monochromatic solutions, other than
$\ket{\psi_\mathrm{out}}=\ket{0^n}$,
namely $\ket{\psi_\mathrm{out}}=\ket{\vec{k}}$, with
$1\le\vec{k}\le{2^{n}-1}$,
if and only if the test function $f$
is either linear or affine, namely $f(\vecx)=\vec{k}\cdot\vecx\oplus{c}$,
with $c\in\booldomain[]$.

The general output from the Deutsch-Jozsa algorithm can be written as the
following linear combination in the $\tensorpower\Hilbert$ Hilbert space,
namely:
\begin{equation}\label{eq:01}
\ket{\psi_\mathtt{out}}
=
\sum_{\vecp\in\booldomain}\psi_f(\vecp)\ket{\vecp}
\end{equation}
where the probability amplitude $\psi_f(\vecp)$ of having the pure
$n$-qubit state $\ket{\vecp}$ as response (measuring) is given by
the following transform \cite{Nielsen-Chuang:2000}
\begin{equation}\label{eq:03}
 \psi_f(\vecp)=\frac1{2^n}
 \sum_{\vecx\in\booldomain}
 (-1)^{f(\vecx)\oplus\vecp\cdot\vecx}
\end{equation}

\subsection{Walsh transform}
\par
Recalling that the Walsh transform of a Boolean function
$f:\booldomain\mapsto \booldomain[]$
is another Boolean function,
$\hat{f}:\booldomain\mapsto\booldomain[],$
which transforms a bit string $\vec p\in\booldomain$ into the Boolean scalar
\begin{equation}\label{eq:04}
 \hat{f}(\vecp) =
 \sum_{\vecx\in\booldomain}
    (-1)^{f(\vecx)\oplus\vecp\cdot\vecx}
\end{equation}

Confronting equations~(\ref{eq:03}) and (\ref{eq:04}) one finds
$\psi_f(\vecp)$ in terms of the Walsh transform as follows
\begin{equation}\label{eq:05}
 \psi_f(\vecp)=\frac1{2^n}\,\hat{f}(\vecp)
\end{equation}

\section{Bent function and its response to the Deutsch--Jozsa algorithm}
A bent function is a very particular case of Boolean function which can be
defined as follows:

\begin{defn}
A bent function is the Boolean function
$f:\booldomain\mapsto\booldomain[]$
whose Walsh transform
$\hat{f}:\booldomain\mapsto\booldomain[]$
has constant absolute value.
\end{defn}

Rothaus \cite{Rothaus:1976} proved that a bent function has the following
metric signature:
\begin{equation}\label{eq:06}
\left|\hat{f}(\vecp)\right|={2^{n/2}},\ \ \
\forall\vecp\in\booldomain
\end{equation}
and that $n$ must be is even.

An example in algebraic normal form of bent function in the two-bit domain
$\{0,1\}^2$ is $f(x_1x_2)=x_1\wedge x_2.$ In this particular,
$\left|\hat{f}(\vecp)\right|=2,\ \
\forall\vecp\in\{0,1\}^2.$

Therefore, if $f$ is a bent function, the spectrum of probability amplitude
$|\psi_f(\vecp)|$ is flat through the $[0,2^n-1]\equiv\booldomain$ integer
interval. Thus, according to equations~(\ref{eq:05}) and (\ref{eq:06}), one
finds that
\begin{equation}\label{eq:09}
\left|\psi_f(\vecp)\right|=\frac{1}{\sqrt{2^n}}
\end{equation}
which is consistent with the normalization principle of quantum mechanics,
namely
\begin{equation}\label{eq:08}
\sum_{\vecp\in\booldomain}
    \left|\psi_f(\vecp)\right|^2
    =1
\end{equation}

\section{Computer simulation of bent functions exercised by the Deutsch--Jozsa
algorithm}

\begin{figure}[h]
 \centering
 \includegraphics[width=6cm,keepaspectratio=true]{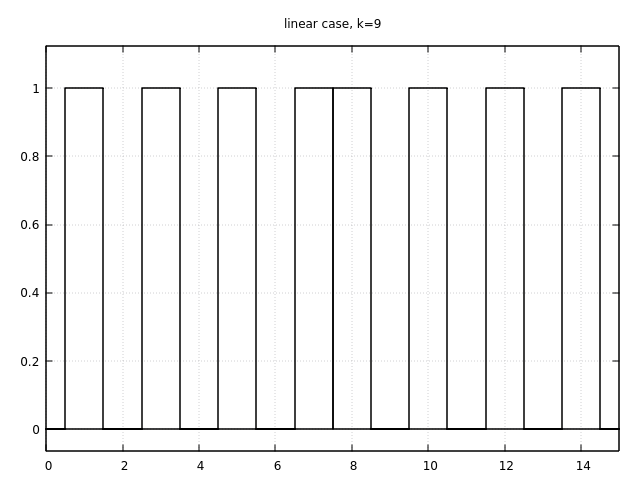}
 \caption{The linear case for k = 9 (see main text)}
 \label{fig:01}
\end{figure}

\begin{figure}[h]
 \centering
 \includegraphics[width=6cm,keepaspectratio=true]{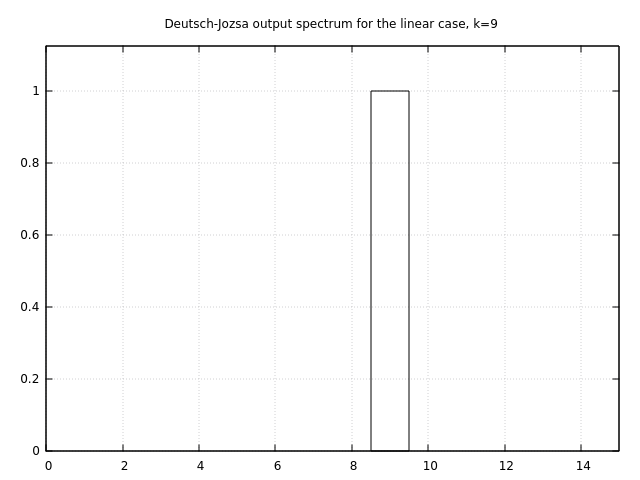}
 \caption{The Deutsch-Jozsa output spectrum for the linear case shown in
previous figure}
 \label{fig:02}
\end{figure}

\par
To illustrate the behavior of the Deutsch-Jozsa algorithm, a C-language program
was implemented to simulate Boolean functions, for linear, arbitrary, no
curvature, simpler curvature, and random curvature. The output of the quantum
algorithm is simulated using Equation (3) to obtain the probability
distribution function as the square of the amplitudes in the equation.

\par
The simulations were made considering only 4 bits, which is enough for a good
visualization. The first simulated case was the linear case, say $f(\vec x) =
\vec k \cdot \vec x$ \cite{Marinho:2018}, with $\vec k = 9$. The result is
shown in Figures 1 and 2, the latter being the simulation of what would be
observed in the output of the Deutsch--Jozsa algorithm after a large number of
runs.

\begin{figure}[h]
 \centering
 \includegraphics[width=6cm,keepaspectratio=true]{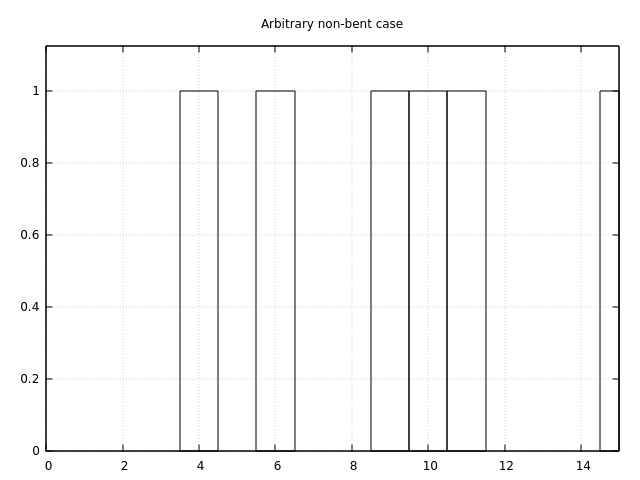}
 \caption{An arbitrary, non-bent Boolean function}
 \label{fig:03}
\end{figure}

\begin{figure}[h]
 \centering
 \includegraphics[width=6cm,keepaspectratio=true]{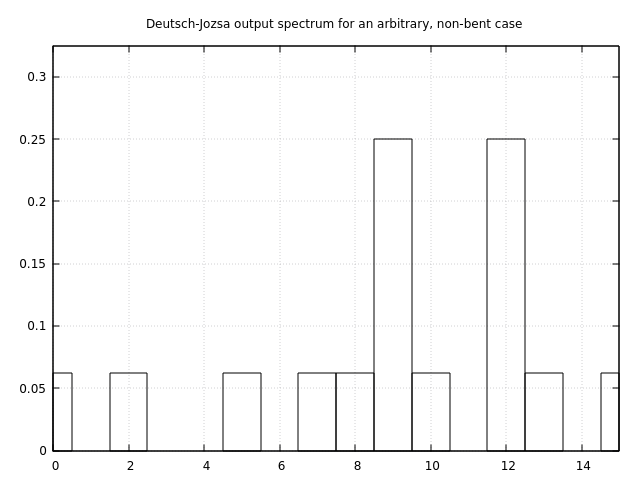}
 \caption{The Deutsch-Jozsa response for the non-bent case shown in the
previous figure}
 \label{fig:04}
\end{figure}

\par
Figures 3 and 4 show the result of a simulation for an arbitrary, non-bent
function. Compare the difference between the output spectra, Figures 2 and 4,
between the linear case and the arbitrary case. Also note that the non-bent
case has an arbitrary probability distribution.

\begin{figure}[h]
 \centering
 \includegraphics[width=6cm,keepaspectratio=true]{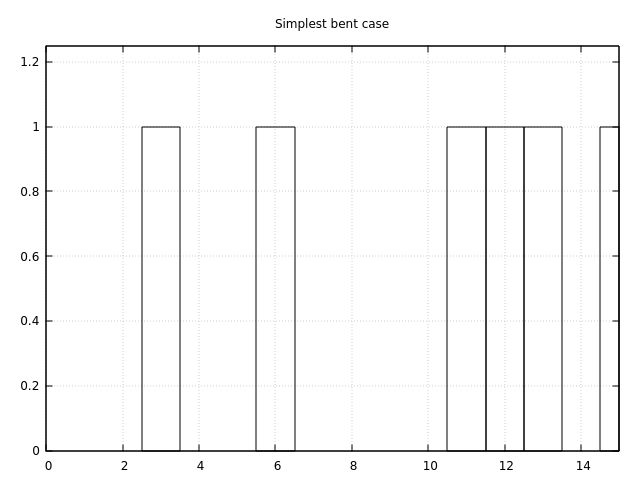}
 \caption{The simplest case of 4-bit bent function (see main text)}
 \label{fig:05}
\end{figure}

\begin{figure}[h]
 \centering
 \includegraphics[width=6cm,keepaspectratio=true]{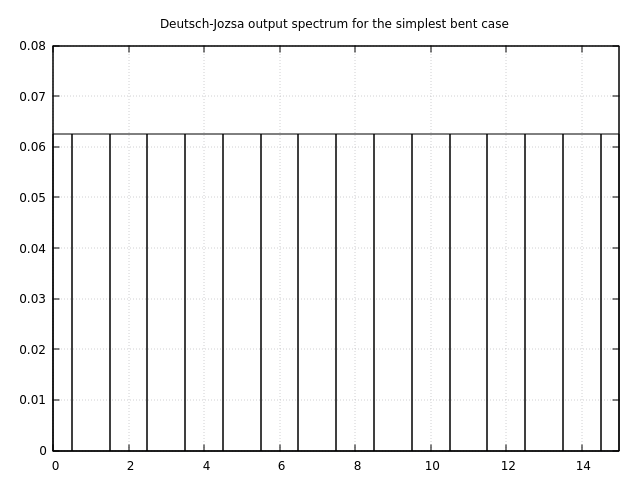}
 \caption{The Deutsch-Jozsa response for the simplest bent function of previous
figure}
 \label{fig:06}
\end{figure}

\par
Figures 5 and 6 illustrate the simplest case of bent function, namely
$f(x_3x_2x_1x_0)=x_0\wedge x_1+x_2\wedge x_3$, where $\wedge$ means the bitwise
\texttt{AND} operator. In this case, the
probability spectrum is flat, as shown in Figure 6.

\begin{figure}[h]
 \centering
 \includegraphics[width=6cm,keepaspectratio=true]{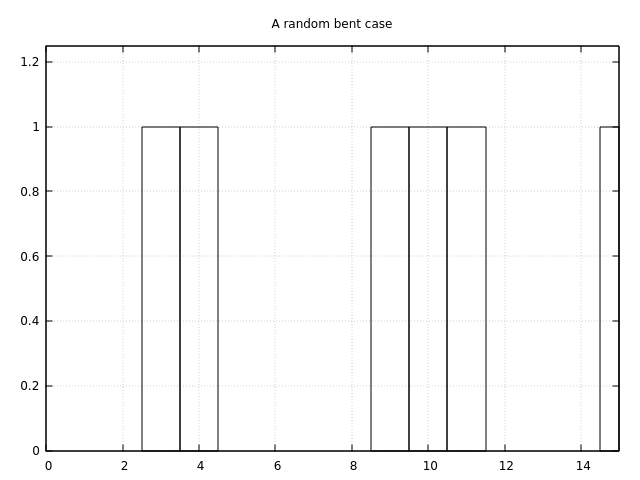}
 \caption{A 4-bit arbitrary bent function}
 \label{fig:07}
\end{figure}

\begin{figure}[h]
 \centering
 \includegraphics[width=6cm,keepaspectratio=true]{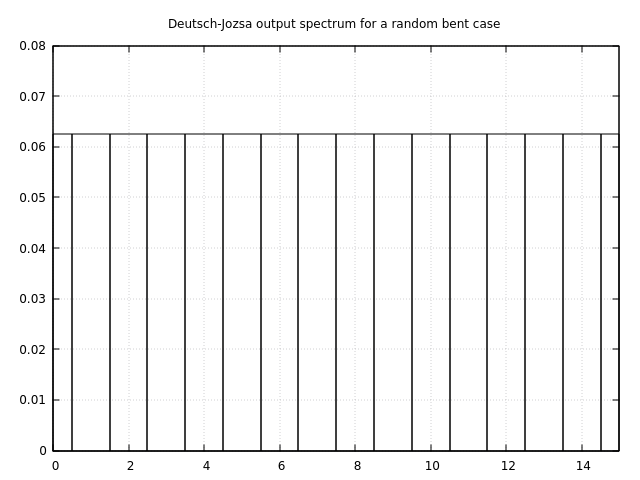}
 \caption{The Deutsch-Jozsa spectrum for the arbitrary bent function shown in
previous figure}
 \label{fig:08}
\end{figure}

\par
Finally, Figures 7 and 8 illustrate a case of randomly chosen bent function. The
program has the option of random shuffling, and after a few runs, the one that
returns a flat probability response is chosen.

\section{Conclusion}
It was shown in the previous section that the Deutsch--Jozsa algorithm responds
with a flat spectrum of solutions if the test function is a bent function.

Particularly important is that the original purpose of the Deutsch-Jozsa
algorithm was to decide whether a Boolean function was constant or balanced,
and now we are showing that the same quantum algorithm works on deciding if a
function is bent or not.


\bibliographystyle{acm}


\begin{thebibliography}{1}

\bibitem{Deutsch-Jozsa:1992}
{\scshape Deutsch, D., and Jozsa, R.}
\newblock Rapid solution of problems by quantum computation.
\newblock {\em Proceedings of the Royal Society of London A: Mathematical,
  Physical and Engineering Sciences 439}, 1907 (1992), 553--558.

\bibitem{Marinho:2018}
{\scshape Marinho, E.~P.}
\newblock Beyond z=0. the deutsch-jozsa decided monochromatic languages.
\newblock {\em arXiv e-prints\/} (Dec. 2018), {1--25. }.

\bibitem{Nielsen-Chuang:2000}
{\scshape Nielsen, M.~A., and Chuang, I.~L.}
\newblock {\em Quantum Computation and Quantum Information}.
\newblock Cambridge University Press, 2000.

\bibitem{Rothaus:1976}
{\scshape Rothaus, O.}
\newblock On “bent” functions.
\newblock {\em Journal of Combinatorial Theory, Series A 20}, 3 (1976), 300 --
  305.

\end{thebibliography}

\end{document}